\documentclass[aps,pra,twocolumn,showpacs,superscriptaddress]{revtex4}

\usepackage{amssymb,amsmath,graphicx,graphics}

\begin{document}

\title{Trapped Bose-Einstein condensates in the presence of a current nonlinearity}
\author{Mohammed F. Saleh}
\affiliation{SUPA, Institute of Photonics and Quantum Sciences, Heriot-Watt University, EH14 4AS Edinburgh, United Kingdom}
\affiliation{Department of Mathematics and Physics Engineering, Alexandria University, Alexandria, Egypt}
\author{Patrik \"{O}hberg}
\affiliation{SUPA, Institute of Photonics and Quantum Sciences, Heriot-Watt University, EH14 4AS Edinburgh, United Kingdom}

\begin{abstract}
We investigate the effect  of a current nonlinearity  on the evolution of a trapped atomic Bose-Einstein condensate. We have implemented techniques from the field of nonlinear optics to provide new insights into the irregular dynamics associated with chiral superfluids. We have found that the current nonlinearity can be treated as a Kerr-like nonlinearity modulated by a spatiotemporal function that can lead to a number of processes such as broadening and compression of the wave function. In the long time scale limit, the wave function is drastically deformed and delocalised compared to its initial state. However, localised modes which oscillate with the period of the inverse trap frequency can still be observed. 
\end{abstract}

\maketitle

\section{Introduction}

At sufficiently low temperatures a gas of bosonic atoms may form a Bose-Einstein condensate (BEC). This state of matter has provided a testbed for a plethora of exotic phenomena at the quantum level. Prominent examples include phenomena related to many-body physics such as, for instance, the superfluid to Mott insulator transition \cite{greiner_2002}, nonlinear dynamics including solitons \cite{burger_1999,denschlag_2000} and quantised vortices \cite{Matthews1999a,Madison2000}, and recently also the creation of  synthetic magnetic fields for the neutral atoms \cite{lin_2009b,Aidelsburger:2011,dalibard_2011,goldman_2013}. 

The advent of synthetic magnetic fields and orbital magnetism in Bose-Einstein condensates has opened up new avenues of research. The neutral gas can now be used for emulating magnetic solid state phenomena, but also more exotic scenarios where the gauge potential in question is non-Abelian which consequently provides links with high energy physics and the Standard Model \cite{goldman_2013}. The synthetic magnetic field can be created using many different techniques, ranging from rotating the system, laser-induced tunneling, Floquet-type dynamical variations of the lattice parameters in optical lattices,  or  inducing a geometric potential from light-matter coupling \cite{goldman_2013}.  These artificial gauge potentials or magnetic fields are, however, static. They are given by external parameters such as the laser intensity or phase gradients of the incident laser field. The resulting gauge field is therefore not dynamical in the sense that it is not possible to emulate completely the Maxwell's equations for instance. In other words, any dynamics of the condensate will not affect the shape of the gauge potential. For this to be the case we need some kind of back-action between the matter field and the gauge potential. This can be achieved using carefully tuned interactions in optical lattices  \cite{banerjee_2012,zohar_2012,tagliacozzo_2013,greschner_2014}, atoms in cavities with strong coupling between the cavity field and the atoms \cite{dong_2014,zheng-cooper_2016,ballantine_2017}, or collisionally induced detunings for geometric potentials which gives rise to nonlinear gauge potentials and current nonlinearities \cite{edmonds_2013a,zheng_2014}. 

In this paper we will consider collisionally induced detunings and the resulting gauge potentials \cite{edmonds_2013a,zheng_2014}. With this technique we do not emulate a fully dynamical gauge theory, but rather, a nonlinear, interacting gauge theory, with unconventional superfluid dynamics. Our goal is to investigate the nonlinear dynamics of a harmonically trapped Bose-Einstein condensate which is subject to a current nonlinearity. We will in particular draw analogies with phenomena and models from nonlinear optics, where similar governing equations are used. The motivation for this is two-fold. Firstly, we want to shed some light on the irregular type of dynamics that occurs in the presence of a current nonlinearity. The analogies between nonlinear optics of ultrashort pulses, and superfluid dynamics in ultracold gases, may well provide important new insight into the onset of chaos in such atomic systems. Secondly, we are interested in novel type of soliton dynamics in a confined setting, where the underlying dynamics is chiral in nature due to the current nonlinearity. 

The paper is organised as follows. In Sec. II we briefly introduce the physical model and the corresponding equation which describes the dynamics of Bose-Einstein condensate. Sec. III discusses the nonlinear dynamics from a nonlinear optics point of view. Possible experimental realisations of the investigated system have been discussed in Sec. IV. Finally, our conclusions are summarised in Sec. V.


\section{The model}
In a series of recent papers \cite{edmonds_2013a,edmonds_2015,zheng_2014,butera_2016}, it has been shown that a nonlinear gauge potential can be created for charge neutral ultracold atoms. This is an example of how to emulate an interacting gauge theory with cold atomic gases such as a Bose-Einstein condensate.  There are many ways to create synthetic gauge potential for atoms \cite{dalibard_2011,goldman_2013}. We are here interested in dynamical gauge potentials such as density-dependent potentials, where a strong interaction between the matter field and the gauge potential takes place. Such a density-dependent gauge potential can be created if we allow for energy shifts of the internal electron levels due to the collisions between the atoms. 

Perhaps the simplest illustration on how a density dependent gauge potential  can be created is based on two-level atoms whose levels are coupled by an incident laser beam, which may have a space-dependent amplitude and phase. If the corresponding Rabi frequency is the dominating energy scale, a perturbative treatment of the collisional energies gives rise to a density dependent gauge potential of the form
\begin{equation}
{\bf A}(r)={\bf A}_0(r)+\nabla \phi \frac{(g_{11}-g_{22})\varrho_0(r)}{8\Omega}
\end{equation}
where ${\bf A}_0$ is the static gauge potential \cite{goldman_2013}, $\phi$ is the phase of the laser which couples the two levels, $g_{11}=4\pi\hbar^2a_{11}/m$ and $g_{22}=4\pi\hbar^2a_{22}/m$ describe the strength of the collisions and include the corresponding scattering lengths $a_{11}$ and $a_{22}$ for collisions between atoms in state $1$ and $2$ respectively, $\Omega$ is the Rabi frequency, and $\varrho_0$ is the density of the cloud. For a more detailed derivation of the synthetic gauge potential and the resulting equation of motion for the atoms we refer the reader to the work by Edmonds et al. \cite{edmonds_2013a}.

In this paper, we consider a tightly confined quasi one-dimensional BEC where the transversal confinement is strong enough to dynamically freeze all the motion in this direction, but such that the collisional dynamics is still three-dimensional. In the longitudal direction the BEC is weakly confined by a harmonic trap. In this situation the transversal degrees of freedoms can be integrated out and we are left with an effective one-dimensional Gross-Pitaevskii equation
\begin{equation}
i\hbar\dfrac{\partial\psi}{\partial t}=\left[ \dfrac{-\hbar^{2}}{2m}\dfrac{\partial^{2}}{\partial x^{2}}+\dfrac{m\omega^{2}}{2}x^{2}+g\left|\psi\right|^{2}-2\alpha I\left(x \right)\right]  \psi, \label{Eq1}
\end{equation}
where $ \psi\left( x,t\right)  $ is the macroscopic wavefunction, $ x $ is the longitudinal coordinate, $ t $ is the time, $ \hbar $ is the reduced Planck's constant, $ \omega $ is the trap frequency in the transverse plane, $ m $ is the boson mass, $ g=(g_{11}+g_{22}+2g_{12})/4 $ is the effective nonlinear coefficient where $g_{12}$ represents collisions between atoms in state 1 and 2,  $
\alpha$ is the current-nonlinearity strength, and $  I\left(x \right)=\frac{\hbar}{m}\mathrm{Im}\left( \psi^{*}\partial_{\xi}\psi\right) $. The diffraction, parabolic potential, the intensity- and the current-nonlinearities are represented by the first four terms on the RHS of Eq. (\ref{Eq1}), respectively. Introducing $ \tau=t/ t_{0} $, $ \xi=x/x_{0} $, $ \sigma =-gt_{0}/\hbar$, $ t_{0}=\dfrac{2m x_{0}^{2}}{\hbar} $, $ x_{0}=\sqrt{\dfrac{\hbar}{m\omega}} $, $ \rho=\dfrac{4\alpha}{\sqrt{\hbar m\omega}} $, Eq. (\ref{Eq1}) can be written as
\begin{equation}
\dfrac{\partial\psi}{\partial \tau}=\left[i \dfrac{\partial^{2}}{\partial \xi^{2}}-i\xi^{2}+i\sigma\left|\psi\right|^{2}+\frac{\rho}{2}\,\mathcal{I}\left( \xi\right)\right]  \psi,\label{Eq2}
\end{equation}
with $  \mathcal{I}\left( \xi\right)  =\psi^{*}\partial_{\xi}\psi- \psi\partial_{\xi}\psi^{*}$. 

\section{Current nonlinearity}
Expressing the wavefunction as $ \psi=f e ^{i\phi}$ with  $ f $ the magnitude and  $ \phi $ the phase,   we obtain $  \mathcal{I}\left( \xi\right)  = 2if^{2}\partial_{\xi}\phi$. Hence, the current-nonlinearity has an impact on the wavefunction evolution only when the phase is space dependent. The spatial frequency, $\partial_{\xi}\phi$, is said to be chirped, if the phase has a nonlinear dependence on $\xi$. Even for a wavefunction with an initial constant phase, this nonlinearity still plays a significant role on the dynamics because of the other terms in the RHS of Eq. (\ref{Eq1}) that lead to the evolution of the phase. An exception is a spatial soliton, which results from a balance between the diffraction-induced broadening and self-focusing effects due to the positive intensity-nonlinearity, settled at the minimum of the parabolic potential. For a soliton with an initial spatial profile, $ \psi\left( \xi,0\right) =N \mathrm{sech}\left(N\xi/\sqrt{2} \right)  $, where $ N $ is  an arbitrary parameter which controls the soliton amplitude and spatial width, the chirp induced via the diffraction and the intensity-nonlinearity are canceled by each other, while the influence of the parabolic potential is zero since the soliton remains at  the potential minimum during the evolution. Other initial conditions will lead to a spatially-dependent phase of the wavefunction phase which will consequently affect the evolution of the current nonlinearity. For initial bright soliton states with sech-profiles and $ \sigma=1 $, the parabolic potential is the only source for this dependency if the wavefunction is displaced from the  potential minimum.  Diffraction and nonlinear effects will also start to affect the dynamics as soon as $\sigma$ deviates from unity. 

Alternatively, $  \mathcal{I}\left( \xi\right)   $ can also be written as $ 2i\left|\psi\right|^{2}\partial_{\xi}\phi $. Hence, the current-nonlinearity can be viewed as another Kerr-like nonlinearity modulated by a spatiotemporal function $\partial_{\xi}\phi$. If this function has positive (negative) values, the wavefunction is localised or compressed (broadened or dispersed) via a focusing (defocusing) Kerr-like nonlinearity.

\begin{figure}
\includegraphics[width=8.6cm]{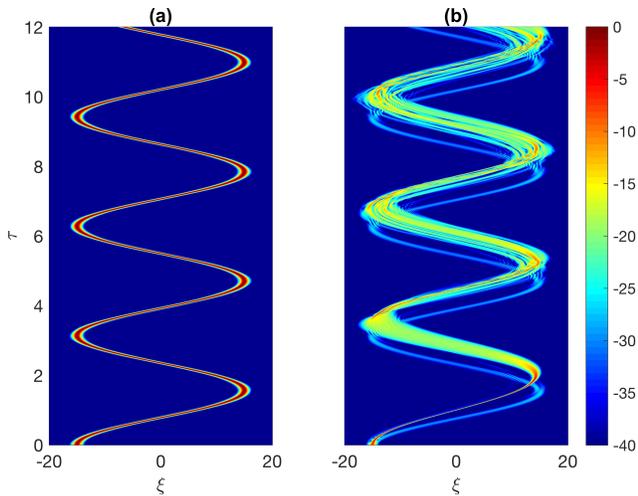}
\caption{Spatial evolution of a wavefunction in the form of $\psi(\xi)= N\mathrm{sech}\left[ N\left( \xi-\xi_{0}\right)/\sqrt{2} \right]  $ in a parabolic potential with $N=5$, $ \xi_{0}=-15 $, and  $ \sigma=1  $  when (a) $ \rho=0 $ and (b) $ \rho=0.4 $. Contour plots are given in a logarithmic scale and truncated at -40 dB. \label{Fig1}}
\end{figure}

\begin{figure}
\includegraphics[width=8.6cm]{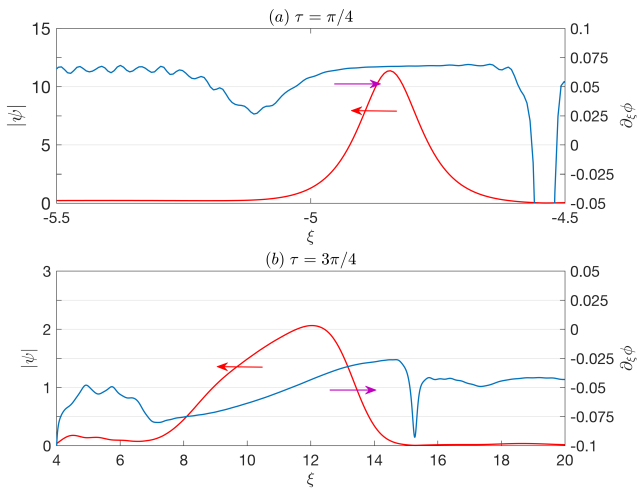}
\caption{Amplitude $\left|\psi\right|  $ and spatial frequency $ \partial_{\xi}\phi $ of the wavefunction at (a) $ \tau=\pi/4 $, and (b) $ \tau=3\pi/4 $ in the presence of the current nonlinearity $ \rho=0.4 $. \label{Fig2}}
\end{figure}

When the wavefunction is displaced from the minimum of the potential, it will start to oscillate  with a period equal to $ \pi $. The effect of the current-nonlinearity on the  oscillation of a wavefunction with an input Sech-profile is depicted in Fig. \ref{Fig1} after simulating Eq. (\ref{Eq2}) using the split-step Fourier method \cite{Agrawal07}, when $ \sigma=1 $.  In this case, the diffraction effects are counteracted by those of the intensity nonlinearity. Panels (a,b) represent the cases when $ \rho=0 $ and $ \rho=0.4 $, respectively.  In the absence  of the current-nonlinearity, the wavefunction shows a periodic harmonic oscillation inside the potential. Contrarily, switching on the current-nonlinearity results in an asymmetric oscillation. The wavefunction is compressed while traveling from left to right, and broadened when moving in the other direction. This is because the wavefunction experiences a positive phase gradient $\partial_{\xi}\phi$ during the first half of the period and a negative  $\partial_{\xi}\phi$ in the second half as shown in Fig. \ref{Fig2}. The regular revival of the dynamics of the wavefunction with a period $\pi$ is therefore broken which can result in a dramatic deformation of the wavefunction compared to the case with no current nonlinearity. This process continues until a pseudo steady state solution is reached which can be characterised by a centre of mass oscillation period which is longer then $\pi$.

Interestingly, an emission of a localised harmonic solution that can maintain the $\pi-$period during propagation is obtained in the simulations presented in Fig. \ref{Fig1}. This soliton-like behaviour is  obtained in the regime when the current nonlinearity is described as a perturbation term in Eq. (\ref{Eq2}), and when the wavefunction is launched in the negative half of the parabolic potential. In this case, the wavefunction is initially prepared with the correct spatial distribution that can overcome the chirp induced by the various sources involved in the dynamics. However, launching the wavefunction on the positive half of the parabolic potential, the wavefunction expands and deforms quickly before it can excite the localised state.

\begin{figure}
\includegraphics[width=8.6cm]{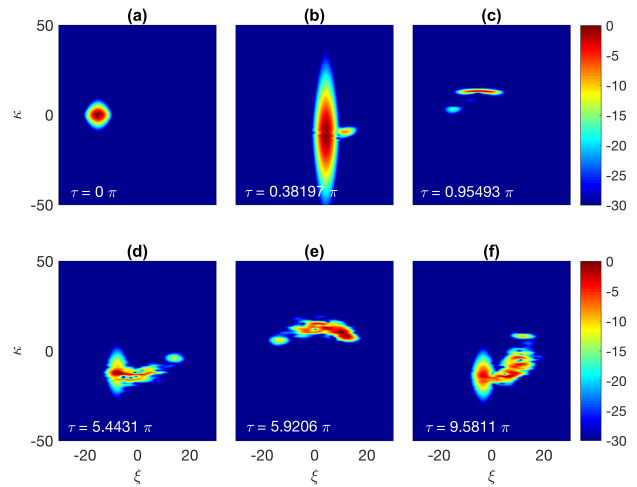}
\caption{ XFROG spectrograms of the wavefunction used in Fig. \ref{Fig1} with $ \sigma=1  $ and $ \rho=0.4 $ at (a) $ \tau=0 $,  (b) $ \tau=0.38\pi $,  (c) $ \tau=0.95\pi $,  (d) $ \tau=5.44\pi $,  (e) $ \tau=5.92\pi $,  (f) $ \tau=9.58\pi $. Contour plots are given in a logarithmic scale and truncated at -30 dB. \label{Fig3}}
\end{figure}

To better visualise the temporal evolution of the  wavefunction inside the parabolic potential in the presence of the current-nonlinearity, we show in Fig. \ref{Fig3} the cross-frequency-resolved-optical-gating (XFROG) spectrogram  at different times during evolution, which is given by \cite{Saleh07}
\begin{equation}
S\left( \xi,\kappa\right)=\left|\displaystyle\int^{\infty} _{-\infty} \psi\left(\xi' \right)  \varphi\left(\xi-\xi' \right) e^{-i\kappa\xi'} d\xi'\right|^{2},
\end{equation}
where  $ \kappa $ is the spatial frequency, and $ \varphi\left(\xi\right)=\exp\left(-\xi^{2} /9\right)  $ acts as a reference gating pulse.  Each spectrogram  displays  the spatial distribution and the corresponding wave vector components on the horizontal and vertical axes, respectively. The input pulse is in the form of a sech-pulse, as  shown in panel (a).  During the first-half of the period, the wavefunction is spatially well localised and the spectrum is broadened, see panel (b).  On the other hand, the wavefunction is broadened in space and localised in spectrum in the second half, as depicted in panel (c). A localised `soliton' solution that preserves its spatial shape during the wavefunction evolution in both directions is also emitted. During the first few periods, the main wavefunction is continuously distorted via spreading in the spatial and spectral domains, as portrayed in panels (d,e). The final deformed pattern of the input wavefunction is shown in panel (f). 
 
\begin{figure}
\includegraphics[width=8.6cm]{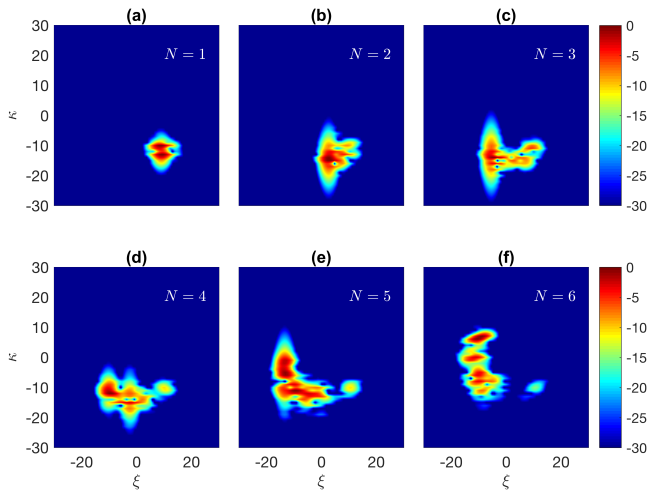}
\caption{ XFROG spectrograms of a wavefunction  inside the parabolic potential in the presence of the current-nonlinearity with  $ \xi_{0}=-15 $,   $ \tau=12  $, $\sigma =1 $ and $ \rho=0.4 $ when (a)  $ N=1  $, (b)  $ N=2  $, (c)  $N=3  $, (d)  $ N=4  $, (e)  $ N=5  $, (f)  $ N=26 $. Contour plots are given in a logarithmic scale and truncated at -30 dB. \label{Fig4}}
\end{figure}

\begin{figure}
\includegraphics[width=8.6cm]{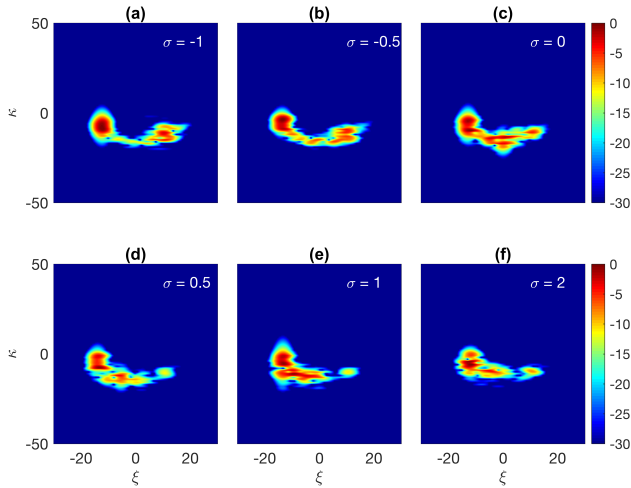}
\caption{ XFROG spectrograms of a wavefunction  inside the parabolic potential in the presence of the current-nonlinearity with $N=5$, $ \xi_{0}=-15 $,   $ \tau=12  $, and $ \rho=0.4 $ when (a)  $ \sigma=-1  $, (b)  $ \sigma=-0.5  $, (c)  $ \sigma=0  $, (d)  $ \sigma=0.5  $, (e)  $ \sigma=1  $, (f)  $ \sigma=2  $. Contour plots are given in a logarithmic scale and truncated at -30 dB. \label{Fig5}}
\end{figure}

In the following, we will investigate how the wavefunction is affected when varying the free parameters, in particular, the wavefunction amplitude $N$, the nonlinear strength $\sigma$, and the current nonlinearity $\rho$. Figure \ref{Fig4} depicts the XFROG spectrograms  for different input amplitudes of the wavefunction at a relatively long evolution time with specific values of  $\sigma$ and $\rho$. It is clear that the wavefunction distortion and expansion scales with $N$. The XFROG spectrograms shown in Fig. \ref{Fig5} picture the effect of changing the sign of the intensity nonlinearity $\sigma $ on the wavefunction after a certain evolution time. The major effect is the generation of a localised state when $\sigma $ is positive.  The deformation of the wavefunction is weakly-dependent on the sign and the magnitude of  $\sigma $.

\begin{figure}
\includegraphics[width=8.6cm]{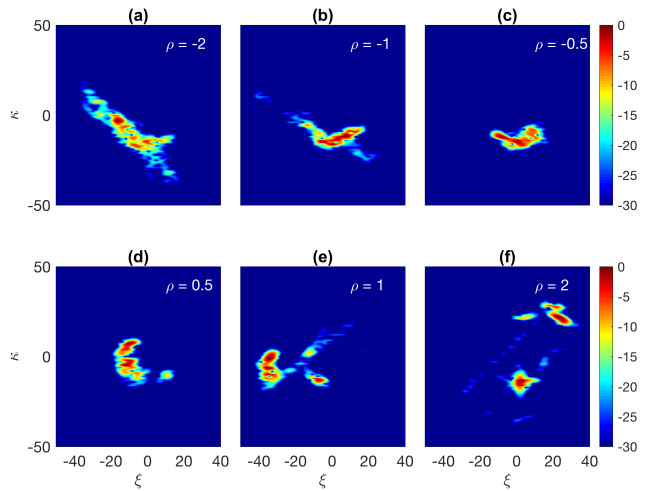}
\caption{ XFROG spectrograms of a wavefunction   inside the parabolic potential in the presence of the current-nonlinearity with $N=5$, $ \xi_{0}=-15 $,   $ \tau=12  $, and $ \sigma=1 $ when (a) $ \rho=-2 $, (b) $ \rho=-1 $, (c) $ \rho=-0.5$, (d) $ \rho=0.5 $, (e) $ \rho=1 $, (f) $ \rho=2 $.  Contour plots are given in a logarithmic scale and truncated at -30 dB.  \label{Fig6}}
\end{figure}

The evolution of the XFROG spectrograms for different current nonlinearity strengths $\rho$ is depicted in Fig. \ref{Fig6}. The spectrograms vary dramatically as $\rho$ is scanned from negative to positive values. At large negative values, the wavefunction diffuses over the spatial and spectral domain. For instance, when $\rho=-2$ the wavefunction is linearly negatively-chirped, as shown in panel (a). As $\rho$ increases, the width of the wavefunction starts to shrink, see panels (b,c). However, the regime characterised by positive values of $\rho$ is distinguished by the emission of the localised modes, portrayed in Fig. \ref{Fig6}(d--f). The wavefunction disintegrates into multiple localised states when $\rho$ increases. We have found that the wavefunction reaches its steady state solution quickly as $\left|\rho\right|$ grows. This demonstrates that the current nonlinearity plays a significant role  in the dynamics of the evolution of the Bose-Einstein condensate.

\section{Experimental realisation}

A number of criteria need to be fulfilled in order to realise a current nonlinearity in the Bose-Einstein condensate. Firstly the density-dependent gauge potential relies on the Rabi frequency and the corresponding energy scale to dominate over any collisional interaction energies. In practice this means $\hbar\Omega$ must be larger than the chemical potential $\mu$ of the Bose-Einstein condensate. Secondly, we need to use  suitable  atomic states and scattering lengths, such that a non-zero current nonlinearity can be obtained. For this one needs the scattering lengths $g_{11}\ne g_{22}$  which, if not readily available, can be achieved using Feshbach resonances. Also, we need to be in the adiabatic regime, where the dressed states arising from the light-matter interaction are not coupled. This not only requires that $\hbar\Omega\gg\mu$, but the atomic states must also be long lived. Promising candidates could for instance be Ytterbium and Strontium which have extremely long lived states of the order of seconds \cite{dalibard_2011}. An alternative route in order to avoid spontaneous emission and heating is to use dark states and three-level atoms \cite{theuer_1999,dalibard_2011}. In the one-dimensional situation discussed in this paper we have chosen a particularly simple form of the phase of the incident laser which is represented by a plane wave. This results in a gauge potential which is proportional to the density of the Bose-Einstein condensate. The phase of the laser can however be shaped quite freely using, for instance, Spatial Light Modulator technology\cite{mcgloin2003applications}, which consequently would also allow for a space dependent $\rho$-parameter controlling the strength of the current-nonlinearity.

\section{Conclusions}
In conclusion, we have studied the evolution of a trapped Bose-Einstein condensate which is subject to a current nonlinearity.  We have adopted the cross-frequency-resolved-optical-gating (XFROG) technique which displays the spatial and spectral components of the wavefunction as a function of time. We have found that this kind of nonlinearity plays a significant role in the associated dynamics when the wave vector or the phase of the wavefunction becomes space dependent. This leads to various processes of compression and expansion of the wavefunction during its evolution due to successive changing of the sign of the current nonlinearity which results in focusing and defocusing effects. The wavefunction  reaches a steady-state solution after having completely deformed and with significant deviations from the $\pi$-period. Under certain conditions, a localised state that is not influenced by the induced chirp is observed in simulations with a $\pi$-period for small values of current nonlinearities. Multiple localised states have also been obtained by increasing the strength of the current  nonlinearity. These rich mechanisms demonstrate the complexity associated with the dynamics of density-dependent gauge potentials, and open new links between ultracold gases and nonlinear optics.

\acknowledgments
The authors acknowledge helpful discussions with Fabio Biancalana and Robert Dingwall.
M.S. would like to acknowledge the support of his research by the Royal Society of Edinburgh and the Scottish Government. P.\"O. acknowledges support from EPSRC grant No. EP/M024636/1.


\end{document}